\documentstyle[pra,aps]{revtex}
\begin{document}
\draft
\title{Test Particle in a Quantum Gas}
\author{Bassano~Vacchini\footnote{E-mail: bassano.vacchini@mi.infn.it}}
\address{Dipartimento di Fisica  
dell'Universit\`a di Milano and Istituto Nazionale di Fisica  
Nucleare, Sezione di Milano, Via Celoria 16, I-20133, Milan,  
Italy}  
\date{\today}
\maketitle
\begin{abstract}
A master equation with a Lindblad structure is derived, which describes
the interaction of a test particle with a macroscopic system
and is expressed in terms of the operator valued dynamic
structure factor of the
system. In the case of a free Fermi or Bose gas the result is
evaluated in the  Brownian limit, thus obtaining a single
generator master equation for the description of  quantum
Brownian motion in which the correction due to quantum
statistics is explicitly calculated.
The friction coefficients for Boltzmann and
Bose or Fermi statistics are compared.
\end{abstract}
%%%%%%%%%%%%%%%%%%%%%%%%%%%%%%%%%%%%%%%%%%%%%%%%%%%%%%%%%%%%%
\section{INTRODUCTION}  
%%%%%%%%%%%%%%%%%%%%%%%%%%%%%%%%%%%%%%%%%%%%%%%%%%%%%%%%%%%%%
The study of the dynamics of a particle coupled to a general
many-body system plays a relevant role in 
modern quantum physics, both with respect to foundations and
applications of  quantum theory.
On the one hand it provides a most simple example of quantum
dynamics of a non isolated system, possibly offering a
manageable arena for a truly microscopic approach, which might
shed some light on mechanisms of dissipation and
decoherence~\cite{Kiefer}: these last two issues are now of
outstanding relevance in connection with the rapidly growing
experimental ability to deal with thoroughly  quantum
mechanical phenomena, checking for their coherence
properties (at the single  particle level think for example
of the recent cavity QED and ion trapping
experiments~\cite{qed-ion}, while at many-body level Bose-Einstein
condensation is a most interesting example~\cite{bec}).
On the other hand plenty of interesting physical problems
may be modeled in this way and among these, in particular,
motion or diffusion of charged or neutral  particles in
gases, liquids or solids.
The interaction of a test particle with a dilute or
noninteracting gas is strictly connected to the problem of a
quantum generalization of the Boltzmann equation, whose
everlasting relevance has recently been stressed by the
experimental realization of quantum degenerate samples of
weakly interacting bosons or fermions~\cite{bec,fermi}: in fact for
the study of these systems resort is often made to a quantum
Boltzmann transport equation~\cite{Imoto-Ferrari-QKI}.
A particularly interesting situation arises if the mass $M$
of the test particle is much bigger than the mass $m$ of the
particles which make up the gas: the so called  Brownian
motion, which serves as a paradigmatic example in the
description of irreversible and dissipative processes. The
description of the phenomenon is still debated at a quantum
level (see~\cite{art3} and references cited therein), even
though well settled by now at classical level in terms of
Langevin or Fokker-Planck equations (it took however 
almost a century from the observation by
Brown to the first successful theoretical description by
Einstein, which led to the first example of
fluctuation-dissipation relation, linking friction and diffusion
coefficients).
A class of models, usually named quantum Brownian 
motion~\cite{LindbladQBM-AlbertoQBM-LindbladJMP}, given by time
evolutions with a non Hamiltonian part 
mapping the algebra of operators at most
bilinear in the operators ${\hat {{\sf x}}}$ and
${\hat {{\sf p}}}$ of the particle into itself, seem to be the most natural
candidate in order to obtain equations of motion analogous to
the classical ones, leading in particular to a friction
force proportional to velocity. On the mathematical side
generators of time evolution semigroups satisfying these
requirements have been fully characterized through the
property of complete positivity, which formally amounts to
the requirement that positivity of the time evolution is
preserved even in presence of coupling without interaction
to another system and leads to a typical expression for the
generators of these semigroups, known as Lindblad
structure~\cite{Lindblad-Kraus-Hellwig-Alicki}.
This has led to a wide literature developing this axiomatic
approach~\cite{Sandulescu-Isar}, together with a large number of more or
less phenomenological models in which similar structures are
obtained, though not always preserving
complete positivity (in this connection
see~\cite{Ambegaokar-Pechukas2-IsarJMP-Tannor-97}).
Though warranting positivity of the statistical operator, 
complete positivity is by
itself no fundamental requirement as recently
stressed~\cite{Pechukas-Piza}, so that despite its extensive use in
many fields of physics, ranging from quantum optics to quantum
communication, the study of the conditions and approximations under
which it emerges from microphysical models is strongly desirable.
\par
In recent work the derivation at a
fundamental level of a completely positive master equation for a Brownian
particle interacting with a free Boltzmann gas has been
given, based on a microphysical model developed for the
description of particle matter interaction~\cite{art1,art3}. 
The Lindblad
equation thus obtained can be written with a
single generator and temperature dependent friction and
diffusion coefficients were determined in terms of 
the scattering cross section.
In this paper we give a major extension of the previous
model, keeping also quantum statistics of the gas into
account. Moreover, before going over to the Brownian limit, in which
the ratio between the masses is much smaller than one, one
sees that the generator of the master equation is
expressed in terms of the dynamic
structure factor of the medium, first
introduced by van Hove~\cite{vanHove}. This turns out to be true
also for an interacting system, thus linking in full
generality the dynamics of the test particle to the
density fluctuations of the system [see Eq.~(\ref{general})]. 
The property of complete positivity is retained in the general case under some 
requirements on the energy dependence of the dynamic
structure factor, which are exactly
fulfilled in the case of Boltzmann particles dealt with in~\cite{art3}.
The Brownian limit is
then considered, thus obtaining the correction at finite
temperature due to quantum statistics to the equation
describing quantum Brownian motion [see Eq.~(\ref{11})]. 
In terms of the fugacity $z$ this correction
takes a remarkably simple form [see Eq.~(\ref{f2})].
\par
The paper is organized as follows: in Sec.~II we 
consider the general structure of the master equation and its
connection to the dynamic
structure factor; in Sec.~III we obtain the correction due to
quantum statistics to the master equation describing quantum Brownian
motion, 
together
with the relationship between the friction coefficients for Boltzmann
or quantum statistics; in  Sec.~IV we  
comment on our results indicating potential future developments.
%%%%%%%%%%%%%%%%%%%%%%%%%%%%%%%%%%%%%%%%%%%%%%%%%%%%%%%%%%%%%
\section{GENERAL STRUCTURE OF THE MASTER
EQUATION IN TERMS OF THE DYNAMIC STRUCTURE FACTOR}
%%%%%%%%%%%%%%%%%%%%%%%%%%%%%%%%%%%%%%%%%%%%%%%%%%%%%%%%%%%%%
Let us briefly recall the structure of the master
equation obtained in~\cite{art1} for the description of the
subdynamics of a particle interacting with a macroscopic
system supposed to be at equilibrium. The result is valid on
a time scale $\tau$ much longer than microphysical collision
time and describes an interaction
through two-particle collisions given by the full T matrix.
The master equation 
is given by:
        \begin{equation}
        \label{1}
        {  
        d {\hat \varrho}  
        \over  
                      dt
        }  
        =
        -
        {i \over \hbar}
        [{\hat {{\sf H}}}_0
        ,
        {\hat \varrho}
        ]
        +
        {1\over\hbar}
        \sum^{}_{{\lambda,\xi}}
        \bigl[
        {\hat {\sf L}}_{\lambda\xi} {\hat {\varrho}}
        {{\hat {\sf L}}{}_{\lambda\xi}^{\scriptscriptstyle  
        \dagger}}  
        -
        {\scriptstyle {1\over 2}}  
        \{
        {  
        {{\hat {\sf L}}{}_{\lambda\xi}^{\scriptscriptstyle \dagger}}  
        {\hat {\sf L}}{}_{\lambda\xi}  
                , {\hat \varrho}  
                }
        \}
        \bigr]
        ,
        \end{equation}
with
        \[
        \langle  
        {k}
        \vert  
        {\hat {\sf L}}_{\lambda\xi}
        \vert  
        {h}  
        \rangle  
        =
        \sqrt{2\varepsilon  \pi_\xi}
        {
        \langle
        \lambda
        \vert
        {
        T{}_{h}^{k}
        }
        \vert
        \xi
        \rangle
        \over
        {{E_k}+{E_{{\lambda}}}-{E_h}-{E}_{\xi} -i\varepsilon}
        }
        ,
        \]
where ${\hat {{\sf H}}}_0$ is the Hamiltonian for the particle and
${\hat \varrho}$ its statistical operator, while
$
{{\varrho}^{\text{m}}}=\sum_\xi \pi_\xi
\vert\xi\rangle\langle\xi\vert
$ is
the statistical operator for matter at equilibrium, $\pi_\xi$ being
the statistical weights related to its spectral decomposition.
The vectors $\vert\lambda\rangle$ and
$\vert\xi\rangle$ are eigenvectors of the macrosystem
Hamiltonian
${H}_{\text{m}}$ with eigenvalues $E_\lambda$ and $E_\xi$
respectively, similarly $\vert k\rangle$ and
$\vert h\rangle$ denote eigenvectors of ${\hat {{\sf H}}}_0$
with eigenvalues $E_k$, and $E_h$.
In writing the equation we have neglected the slow energy
dependence of the T matrix, which would have brought a
commutator term proportional to the forward scattering
amplitude, diagonal in momentum representation.
The terms other than the commutator in (\ref{1}) are linked to the
dissipative behavior, which cannot be obtained in a Hamiltonian
formalism. Interactions at microphysical level are typically
translationally invariant, so that a general Ansatz for the T matrix
describing two-body interactions is given by
$
        T{}_{h}^{k}
        =
        {\int d^3 \! {\bbox{x}} \,}
        {\int d^3 \! {\bbox{y}} \,}  
        \psi^{\scriptscriptstyle\dagger}({\bbox{x}})  
        u_k^{*}({\bbox{y}})  
        t({\bbox{x}}-{\bbox{y}})  
        u_h({\bbox{y}})  
        \psi({\bbox{x}})  
$,
where $\psi^{\scriptscriptstyle\dagger}$,
$\psi$ are field  operators for the macrosystem.
We now consider a homogeneous system, so as to use as quantum numbers
momentum eigenvalues, thus obtaining with a Fourier transform an
expression depending only on the modulus of the momentum transfer:
        \begin{equation}
        \label{1bis}
        T{}_{h}^{k} 
        =  
        \sum_{\eta\mu}
        \delta_{p_\eta +p_k,p_h+p_\mu}
        \tilde{t} (
        |
        {\bbox{p}}_{\mu}-{\bbox{p}}_{\eta}
        |
        )
        b^{\scriptscriptstyle\dagger}_{\eta}
        b_\mu                      ,
        \end{equation}
where $b^{\scriptscriptstyle\dagger}$, $b$ denote
creation and destruction operators
in the Fock space of the macrosystem.
Restricting to the case of a free gas of Bose or Fermi
particles, the eigenvectors of ${H}_{\text{m}}$ can be
characterized as a set of occupation numbers $n_\sigma$
relative to particles with a given momentum
${\bbox{p}}_{\sigma}$, so that
$
        \vert\xi\rangle
        =
        \vert
        \{
        n_{\sigma}^\xi
        \}         \rangle
$,
and the matrix element
$
        \langle\lambda\vert
        b^{\scriptscriptstyle\dagger}_\eta b_\mu
        \vert\xi\rangle
$
can be readily evaluated restricted to the primed sum for
$\lambda\neq\xi$, since in the case $\lambda=\xi$ the
contributions to the master
equation  (\ref{1}) cancel out.
Denoting by
$
{{\bbox{q}}}={\bbox{p}}_{\mu}-{\bbox{p}}_{\eta}
$
the  momentum transferred to the test particle and by
${\Delta E}_{\mu q}({\bbox{p}}) =
        {({{\bbox{p}} + {\bbox{q}}})^2\over 2M}
        +
        {({{\bbox{p}}_{\mu} - {\bbox{q}}})^2\over 2m}
        -
        {{{\bbox{p}}}^2 \over 2M}
        -
        {{\bbox{p}}_{\mu}^2\over 2m}
$
the difference in energy before and after the collision ($M$
being the mass of the test particle with momentum
${\bbox{p}}$, $m$ the mass of the gas particles), and
supposing the statistical operator ${\hat \varrho}$ to be
quasi-diagonal in momentum representation, 
according to its slow variability, one
sees that (\ref{1}) for a free test particle
reduces to~\cite{art3}
        \begin{eqnarray}
        \label{2}
        {  
        d {\hat \varrho}  
        \over  
                      dt
        }  
        =
        &-&
        {i \over \hbar}
         \left[  
        {
        {\hat {{\sf p}}}^2
        \over
        2M
        }
        ,
        {\hat \varrho}
        \right]
        \\
        \nonumber
        &+&
        {2\pi \over\hbar}
        \sum_{q}{}'
        {
        | \tilde{t} (q) |^2
        }
        \Biggl[
        \sum_{p p'}
        \sum_{\mu}
        \langle
        n_\mu
        \rangle
        (1\pm
        \langle
        n_{\mu - q}
        \rangle
        )
        \delta    %\!\! 
        \left(%\! 
        {\Delta E}_{\mu q}
        \left(%\! 
        {
        {\bbox{p}}+{\bbox{p}}' \over 2
        }
        %\! 
        \right)
        %\! 
        \right)
        e^{{i\over\hbar}{\bbox{q}}\cdot{\hat {{\sf x}}}}
        \vert {\bbox{p}} \rangle
        \langle {\bbox{p}} \vert
        {\hat \varrho}
        \vert {\bbox{p}}' \rangle
        \langle {\bbox{p}}' \vert
        e^{-{i\over\hbar}{\bbox{q}}\cdot{\hat {{\sf x}}}}
        \nonumber
        \\
        &&
        \hphantom{
        {  
        d {\hat \varrho}  
        \over  
                      dt
        }  
        =
        +
        {2\pi \over\hbar}
        \sum_{q}{}'
        {
        | \tilde{t} (q) |^2
        }
        }
        - {1\over 2}
        \sum_{p}
        \sum_{\mu}
        \langle
        n_\mu
        \rangle
        (1\pm
        \langle
        n_{\mu - q}
        \rangle
        )
        \delta    %\!
        \left(      
        {\Delta E}_{\mu q}({
        {\bbox{p}}
        })
        \right)
        \left \{
        \vert {\bbox{p}} \rangle
        \langle {\bbox{p}} \vert,
        {\hat \varrho}
        \right \}
        \Biggr]
        \nonumber
        \end{eqnarray}
where the $+,-$ signs refer to Bose, Fermi statistics
respectively and
\begin{displaymath}
\langle  n_\mu \rangle
=
{
z
e^{
-
\beta
{
{\bbox{p}}_{\mu}^2
\over
2m
}}
\over
1\mp 
z
e^{
-
\beta
{
{\bbox{p}}_{\mu}^2
\over
2m
}}
}
\end{displaymath}
accordingly,
$z$ denoting the fugacity, determined by the requirement
$
        \sum_{\mu}
        \langle
        n_\mu
        \rangle
=N
$, and
$\beta=1/(k_{\rm {B}}T)$ the inverse temperature.
It is worthwhile introducing the more compact notation
        \begin{equation}
        \label{3}
        S_{\rm \scriptscriptstyle B/F}({\bbox{q}},{\bbox{p}})
        =
        \frac 1n
        \int %\!\!
        {
        d^3 \!
        {\bbox{p}_{\mu}}
        \over
        (2\pi \hbar)^3
        }
        \,  
        \langle
        n_{p_{\mu}}
        \rangle
        (1\pm
        \langle
        n_{{p_{\mu}} - q}
        \rangle
        )
        \delta       %\!\!
        \left(
        {\Delta E}_{{p_{\mu}} q}({
        {\bbox{p}}
        })
        \right)
        \end{equation}
where $n$ denotes the density of particles in the gas and
the function
$
        S_{\rm \scriptscriptstyle B/F}
$
is in fact positive definite. 
The integral in (\ref{3}) can be explicitly calculated
both for bosons and fermions giving at finite
temperature the result
        \begin{eqnarray}
        \label{7}
        S_{\rm \scriptscriptstyle B/F}({\bbox{q}},{\bbox{p}})
        =
        {
        1
        \over
         (2\pi\hbar)^3
        }
        {
        2\pi m^2
        \over
        n\beta q
        }
        {
        \mp
        \over
        1-
        \exp{
        \left[  
        {
        \beta
        \over
             2m
        }
        \left(
        2\sigma({\bbox{q}},{\bbox{p}})q -q^2
        \right)
        \right]
        }
        }
        \log
        \left[
        {
        1\mp z
        \exp{
        \left[  
        -{
        \beta
        \over
             2m
        }
        \sigma^2({\bbox{q}},{\bbox{p}})
        \right]
        }
        \over
        1\mp z
        \exp{
        \left[  
        -{
        \beta
        \over
             2m
        }
        (\sigma({\bbox{q}},{\bbox{p}}) - q)^2
        \right]
        }
        }
        \right],
        \end{eqnarray}
where $\sigma({\bbox{q}},{\bbox{p}})=
\frac{1}{2q}
\left[
(1+\alpha)q^2+2\alpha (
        {\bbox{p}}
        \cdot
        {\bbox{q}}
)
\right]$ is expressed in terms
of the dimensionless variable $\alpha=m/M$, giving the ratio
between the masses.
Expression  (\ref{7}) is exactly the dynamic
structure factor for
a free Bose or Fermi gas at finite temperature, as
one could also directly realize from (\ref{3})~\cite{Lovesey} or from the
equivalent expression in terms of  momentum transfer ${\bbox{q}}$ and
energy transfer $
E=
{
q^2
\over
   2M
}
+
{
        {\bbox{p}}
        \cdot
        {\bbox{q}}
\over
M
}
$
(note that we use as variables
momentum and energy transferred to the particle)
\begin{displaymath}
        S_{\rm \scriptscriptstyle B/F}(q,E)
        =
        {
        1
        \over
         (2\pi\hbar)^3
        }
        {
        2\pi m^2
        \over
        n\beta q
        }
        {
        \mp
        \over
        1-
        e^{\beta E}
        }
        \log
        \left[
        {
        1\mp z
        \exp{
        \left[
        -{
        \beta
        \over
             8m
        }
        {
        (2mE + q^2)^2
        \over
                  q^2
        }
        \right]
        }
        \over
        1\mp z
        \exp{
        \left[
        -{
        \beta
        \over
             8m
        }
        {
        (2mE - q^2)^2
        \over
                  q^2
        }
        \right]
        }
        }
        \right],
\end{displaymath}
where the dependence on the transferred momentum is in 
this case actually only
through the modulus.
In the following we will use, according to convenience, both notations
$S({\bbox{q}},{\bbox{p}})$ and $S({\bbox{q}},E)$, where 
$E\equiv
\Delta E_q ({\bbox{p}})=
E_{p+q} - E_{p}
=
{
q^2
\over
   2M
}
+
{
        {\bbox{p}}
        \cdot
        {\bbox{q}}
\over
M
}
$
is the energy transfer.
The dynamic
structure factor is an important physical quantity of direct
experimental access, essentially
depending on the  statistical properties of the macrosystem
and the kinematics of the collision, appearing in the 
expression of the inelastic
differential cross section for a particle interacting with a
macroscopic sample. 
The relation between differential
cross section and dynamic
structure factor  was
first
derived by van Hove in the case of neutron
scattering~\cite{vanHove} and for scattering from state 
${\bbox{p}}$ to state ${\bbox{p}}' =
{\bbox{p}}+{\bbox{q}}$ is given by
\begin{equation}
  \label{diff}
  \frac{d^2 \sigma}{d\Omega dE}=
\left(\frac{M}{2\pi\hbar^2}\right)^2
\frac{p'}{p}
        {
        | \tilde{t} (q) |^2
        }
  S ({\bbox{q}},E)
.
\end{equation}
The dynamic
structure factor is
expressed in the general case 
as Fourier transform of
the time dependent pair correlation function 
with respect to energy and  momentum transfer, according to 
        \begin{equation}
        \label{new}
        S ({\bbox{q}},E)=
        {  
        1  
        \over  
         2\pi\hbar
        }  
        {
        1
        \over
         N
        }
        \int dt 
        {\int d^3 \! {\bbox{x}} \,}        
        e^{
        {
        i
        \over
         \hbar
        }
        (E t -
        {\bbox{q}}\cdot{\bbox{x}})
        }  
        {\int d^3 \! {\bbox{y}} \,}
        \left \langle  
         N({\bbox{y}})  
         N({\bbox{x}}+{\bbox{y}},t)  
        \right \rangle, 
\end{equation}
where
$
        N({\bbox{y}})
$
denotes the local particle density for the macroscopic
system and
$
        \left \langle  
\ldots
        \right \rangle
$
the ensemble average.
Alternatively the dynamic
structure factor may be written in terms of the Fourier transform 
of the density operator $N({\bbox{y}})$, given by
\begin{equation}
\label{rq}
  \rho_q =         
{\int d^3 \! {\bbox{y}} \,}        
        e^{
        -{
        i
        \over
         \hbar
        }
                {\bbox{q}}\cdot{\bbox{y}}
        }  
         N({\bbox{y}}) 
=
        \sum_{\mu}
        b^{\scriptscriptstyle\dagger}_{\mu}
        b_{\mu+q},
\end{equation}
thus obtaining
\begin{equation}
  \label{qq}
        S ({\bbox{q}},E)=
        {  
        1  
        \over  
         2\pi\hbar
        }  
        {
        1
        \over
         N
        }
        \int dt \, 
        e^{
        {
        i
        \over
         \hbar
        }
        E t 
        }  
         \langle  
          \rho_q^{\scriptscriptstyle\dagger} \rho_q (t)
         \rangle.   
\end{equation}
Expression (\ref{qq}) through the relation (\ref{diff}) allows a
determination of the equilibrium fluctuations of the system in terms of
scattering experiments~\cite{Lovesey} (think for example of the very
interesting applications in the case of neutron scattering
from different states and isotopes of Helium~\cite{Glyde}).
Coming back to (\ref{7}) we note that in the limit of very
small fugacity $z\ll 1$ one recovers 
the result for Maxwell Boltzmann particles
        \begin{equation}
        \label{8}
        S_{\rm \scriptscriptstyle MB}({\bbox{q}},{\bbox{p}})
        =
        {
        1
        \over
         (2\pi\hbar)^3
        }
        {
        2\pi m^2
        \over
        n\beta q
        }
        z
        \exp{
        \left[  
        -{
        \beta
        \over
             2m
        }
        \sigma^2({\bbox{q}},{\bbox{p}})
        \right]
        },
        \end{equation}
which in terms of momentum and energy transfer may also be written
\begin{displaymath}
        S_{\rm \scriptscriptstyle MB}(q,E)
        =
        {
        1
        \over
         (2\pi\hbar)^3
        }
        {
        2\pi m^2
        \over
        n\beta q
        }
        z
        \exp\left[
        -{
        \beta
        \over
             8m
        }
        {
        (2mE + q^2)^2
        \over
                  q^2
        }
        \right]
        .
\end{displaymath}
Recalling expression (\ref{3}) for the dynamic
structure factor one immediately realizes
that the master equation given in  (\ref{2}) can be written in terms
of the dynamic structure factor and
exactly exhibits a 
Lindblad structure provided the dynamic
structure factor evaluated at the arithmetic mean
of ${\bbox{p}}$ and ${\bbox{p}}'$ equals the geometric mean of its
values at the two points. This identity holds true without
approximations in the case of expression (\ref{9}) for a Boltzmann gas 
in the Brownian limit considered in~\cite{art3}. In the general case
this factorization relies on an approximation linked to the
quasi-diagonality of the statistical operator. Keeping the linear
relation between $E$ and ${\bbox{p}}$ into account, the approximation
necessary in order to retain complete positivity 
can be most meaningfully written
        \begin{equation}
        \label{x}
        S_{\rm \scriptscriptstyle}\left({\bbox{q}},
        {
        E+E' \over 2
        }\right)
        \approx
        \sqrt{
        S_{\rm \scriptscriptstyle}({\bbox{q}},E)
        }
        \sqrt{
        S_{\rm \scriptscriptstyle}({\bbox{q}},E')
        }
        \end{equation}
and will depend on the smoothness of the energy dependence of $S$ in
the relevant energy region (note that the neglected terms are at least
quadratic in the energy difference). Exploiting (\ref{x}) 
Eq.~(\ref{2}) can be cast in the following Lindblad structure
granting positivity of the time evolution
        \begin{eqnarray}
        \label{ss}
        {  
        d {\hat \varrho}  
        \over  
                      dt
        }  
        =
        &-&
        {i \over \hbar}
        \left[{
        {\hat {{\sf p}}}^2
        \over
        2M
        }
        ,
        {\hat \varrho}
        \right]
        \\
        &+&
        {2\pi \over\hbar}
        (2\pi\hbar)^3
        n
        \int d^3\!
        {\bbox{q}}
        \,  
        {
        | \tilde{t} (q) |^2
        }
        \Biggl[
        e^{{i\over\hbar}{\bbox{q}}\cdot{\hat {{\sf x}}}}
        \sqrt{
        S_{\rm \scriptscriptstyle B/F}({\bbox{q}},{\hat {{\sf p}}})
        }
        {\hat \varrho}
        \sqrt{
        S_{\rm \scriptscriptstyle B/F}({\bbox{q}},{\hat {{\sf p}}})
        }
        e^{-{i\over\hbar}{\bbox{q}}\cdot{\hat {{\sf x}}}}
        -
        \frac 12
        \left \{
        S_{\rm \scriptscriptstyle B/F}({\bbox{q}},{\hat {{\sf p}}}),
        {\hat \varrho}
        \right \}
        \Biggr],
        \nonumber
        \end{eqnarray}
which may be also written in a more manifest Lindblad form
        \begin{eqnarray}
        \label{ll}
        {  
        d {\hat \varrho}  
        \over  
                      dt
        }  
        =
        &-&
        {i \over \hbar}
        [{\hat {{\sf H}}}_0
        ,
        {\hat \varrho}
        ]
        \\
        &+&
        {2\pi \over\hbar}
        (2\pi\hbar)^3
        n
        \int d^3\!
        {\bbox{q}}
        \,  
        {
        | \tilde{t} (q) |^2
        }
        \Biggl[
        L_{\rm \scriptscriptstyle B/F}({\bbox{q}},{\hat {{\sf p}}},{\hat
        {{\sf x}}})
        {\hat \varrho}
        L_{\rm \scriptscriptstyle B/F}^{\scriptscriptstyle\dagger}
        ({\bbox{q}},{\hat {{\sf p}}},{\hat {{\sf x}}})
        -
        \frac 12
        \left \{
        L_{\rm \scriptscriptstyle B/F}^{\scriptscriptstyle\dagger}
        ({\bbox{q}},{\hat {{\sf p}}},{\hat {{\sf x}}})
        L_{\rm \scriptscriptstyle B/F}({\bbox{q}},{\hat {{\sf p}}},{\hat
        {{\sf x}}}),
        {\hat \varrho}
        \right \}
        \Biggr],
        \nonumber
        \end{eqnarray}
introducing the following generator depending on the operators
${\hat {{\sf x}}}$ and
${\hat {{\sf p}}}$
        \begin{equation}
        \label{6}
        L_{\rm \scriptscriptstyle B/F}({\bbox{q}},{\hat {{\sf p}}},{\hat
        {{\sf x}}})
        =
        e^{{i\over\hbar}{\bbox{q}}\cdot{\hat {{\sf x}}}}
        \sqrt{
        S_{\rm \scriptscriptstyle B/F}({\bbox{q}},{\hat {{\sf p}}})
        }.
        \end{equation}
This is a remarkably simple result since $L_{\rm \scriptscriptstyle B/F}$ 
only
depends on the generator of translations in momentum space
and 
the operator valued dynamic
structure factor. 
Let us note that equation 
(\ref{ss}) or equivalently (\ref{ll}) 
is invariant under translation and rotation and
in particular a statistical operator of the canonical form
$
{\hat \varrho} \propto e^{-
\beta
\frac{{\hat {\sf p}}^2}{2M}
}
$
is a stationary solution.
If instead of a free gas one considers a more general medium
characterized by a dynamic
structure factor $S({\bbox{q}},{\bbox{p}})$, provided the
interaction between particle and medium still satisfies translation
invariance as in (\ref{1bis}) and an approximation of the form
(\ref{x}) holds, the master equation (\ref{1}) still has the form
(\ref{ss}) or equivalently (\ref{ll}) with 
$L({\bbox{q}},{\hat {{\sf p}}},{\hat
        {{\sf x}}})$ 
given by
\begin{displaymath}
          L({\bbox{q}},{\hat {{\sf p}}},{\hat
        {{\sf x}}})
        =
        e^{{i\over\hbar}{\bbox{q}}\cdot{\hat {{\sf x}}}}
        \sqrt{
        S({\bbox{q}},{\hat {{\sf p}}})
        }
\end{displaymath}
and therefore retains a completely positive structure.
To prove this we go back to (\ref{1}), which in the case of a
homogeneous system using (\ref{1bis}) can be written
\begin{equation}
  \label{xx}
        {  
        d {\hat \varrho}  
        \over  
                      dt
        }  
        =
        -
        {i \over \hbar}
        [{\hat {{\sf H}}}_0
        ,
        {\hat \varrho}
        ]
        +
        {\cal L} (\varrho)
\end{equation}
with
\begin{eqnarray}
\label{pra1}
  {\cal L} (\varrho)=
&&
\frac{2\varepsilon}{\hbar}
        \sum^{}_{{\lambda\xi}}
        \sum^{}_{kf}
        \sum^{}_{hg}
\vert {\bbox{p}}_f \rangle
\frac
{
        \sum_{\eta\mu}
        \delta_{p_\eta +p_f,p_k+p_\mu}
        \tilde{t} (
        |
        {\bbox{p}}_{\mu}-{\bbox{p}}_{\eta}
        |
        )
        \langle
        \lambda
        \vert
        b^{\scriptscriptstyle\dagger}_{\eta}
        b_\mu  
        \vert
        \xi
        \rangle
}
{
{{E_k}-{E_f}+{E}_{\xi}-{E_{{\lambda}}} +i\varepsilon}
}
\nonumber
\\
&&
\hphantom{\frac{2\varepsilon}{\hbar}
        \sum^{}_{{\lambda\xi}}
        \sum^{}_{kf}
        \sum^{}_{hg}}
\times
        \langle {\bbox{p}}_k \vert
        {\hat \varrho}
        \vert {\bbox{p}}_h \rangle
\pi_\xi
\frac
{
        \sum_{\eta'\mu'}
        \delta_{p_{\eta'} +p_g,p_h+p_{\mu'}}
        \tilde{t}^* (
        |
        {\bbox{p}}_{\mu'}-{\bbox{p}}_{\eta'}
        |
        )
        \langle
        \xi
        \vert
        b^{\scriptscriptstyle\dagger}_{\mu'}
        b_{\eta'}  
        \vert
        \lambda
        \rangle
}
{
{{E_h}-{E_g}+{E_{{\xi}}}-{E}_{\lambda} -i\varepsilon}
}
\nonumber
\\
&&
-\frac{\varepsilon}{\hbar}
        \sum^{}_{{\lambda\xi}}
        \sum^{}_{k}
        \sum^{}_{fg}
\{
\vert {\bbox{p}}_f \rangle         
\langle {\bbox{p}}_g \vert
,
\varrho
\}
\frac
{
        \sum_{\eta\mu}
        \delta_{p_\eta +p_k,p_g+p_\mu}
        \tilde{t} (
        |
        {\bbox{p}}_{\mu}-{\bbox{p}}_{\eta}
        |
        )
        \langle
        \lambda
        \vert
        b^{\scriptscriptstyle\dagger}_{\eta}
        b_\mu  
        \vert
        \xi
        \rangle
}
{
{{E_f}-{E_k}+{E}_{\xi}-{E}_{\lambda} -i\varepsilon}
}
\nonumber
\\
&&
\hphantom{\frac{2\varepsilon}{\hbar}
        \sum^{}_{{\lambda\xi}}
        \sum^{}_{kf}
        \sum^{}_{hg}}
\times
\pi_\xi
\frac
{
        \sum_{\eta'\mu'}
        \delta_{p_{\eta'} +p_k,p_f+p_{\mu'}}
        \tilde{t}^* (
        |
        {\bbox{p}}_{\mu'}-{\bbox{p}}_{\eta'}
        |
        )
        \langle
        \xi
        \vert
        b^{\scriptscriptstyle\dagger}_{\mu'}
        b_{\eta'}  
        \vert
        \lambda
        \rangle
}
{
{{E_g}-{E_k}+{E}_{\xi}-{E}_{\lambda} +i\varepsilon}
}.
\end{eqnarray}
We now introduce the momentum transfer
$
{{\bbox{q}}}={\bbox{p}}_{\mu}-{\bbox{p}}_{\eta}
$,
$
{{\bbox{q}}}'={\bbox{p}}_{\mu'}-{\bbox{p}}_{\eta'}
$
and 
the Fourier transform $\rho_q$
of the density operator given by (\ref{rq}), so that relabelling the
indexes (\ref{pra1}) becomes
\begin{eqnarray}
\label{pra2}
  {\cal L} (\varrho)=
&&
\frac{2\varepsilon}{\hbar}
        \sum^{}_{{\lambda\xi}}
        \sum^{}_{pp'}
        \sum^{}_{qq'}{}'
        \tilde{t} (q) \tilde{t}^* (q') 
        e^{{i\over\hbar}{\bbox{q}}\cdot{\hat {{\sf x}}}}
        \vert {\bbox{p}} \rangle
        \langle {\bbox{p}} \vert
        {\hat \varrho}
        \vert {\bbox{p}}' \rangle
        \langle {\bbox{p}}' \vert
        e^{-{i\over\hbar}{\bbox{q}}'\cdot{\hat {{\sf x}}}}
\\
&&
\nonumber
\hphantom{\frac{2\varepsilon}{\hbar}
        \sum^{}_{{\lambda\xi}}
}
\times
\frac
{
1
}
{
{{E_p}-{E_{p+q}}+{E}_{\xi}-{E_{{\lambda}}} +i\varepsilon}
}
\frac
{
1
}
{
{{E_{p'}}-{E_{p'+q'}}+{E}_{\xi}-{E_{{\lambda}}}-i\varepsilon}
}
\langle\lambda\vert\rho_q\vert\xi\rangle
\pi_\xi 
\langle\xi\vert\rho_{q'}^{\scriptscriptstyle\dagger}\vert\lambda\rangle
\\
&&
\nonumber
-\frac{\varepsilon}{\hbar}
        \sum^{}_{{\lambda\xi}}
        \sum^{}_{p}
        \sum^{}_{qq'}{}'
        \tilde{t} (q) \tilde{t}^* (q') 
\{
\vert {\bbox{p}} \rangle         
\langle {\bbox{p}} \vert
,
\varrho
\}
\\
&&
\nonumber
\hphantom{\frac{2\varepsilon}{\hbar}
        \sum^{}_{{\lambda\xi}}
}
\times
\frac
{
1
}
{
{{E_p}-{E_{p+q}}+{E}_{\xi}-{E_{{\lambda}}} -i\varepsilon}
}
\frac
{
1
}
{
{{E_{p}}-{E_{p+q'}}+{E}_{\xi}-{E_{{\lambda}}}+i\varepsilon}
}
\langle\lambda\vert\rho_q\vert\xi\rangle
\pi_\xi 
\langle\xi\vert\rho_{q'}^{\scriptscriptstyle\dagger}\vert\lambda\rangle
,
\end{eqnarray}
where the primed sum over ${\bbox{q}}$ and ${\bbox{q}}'$ means that
the contribution for ${\bbox{q}}={\bbox{q}}'=0$ is left out, since in
this case the two terms cancel out.
To proceed further we express the denominators in terms of a Laplace
transform, according to
\begin{displaymath}
(a\pm i\varepsilon)^{-1}=\mp\frac{i}{\hbar}
\int_0^\infty d\tau e^{\pm\frac{i}{\hbar}  (a\pm
  i\varepsilon)\tau}  ,
\end{displaymath}
thus obtaining
\begin{eqnarray*}
  {\cal L} (\varrho)=
&&
\frac{2\varepsilon}{\hbar}
         \sum^{}_{pp'}
        \sum^{}_{qq'}{}'
        \tilde{t} (q) \tilde{t}^* (q') 
        e^{{i\over\hbar}{\bbox{q}}\cdot{\hat {{\sf x}}}}
        \vert {\bbox{p}} \rangle
        \langle {\bbox{p}} \vert
        {\hat \varrho}
        \vert {\bbox{p}}' \rangle
        \langle {\bbox{p}}' \vert
        e^{-{i\over\hbar}{\bbox{q}}'\cdot{\hat {{\sf x}}}}
\\
&&
\hphantom{\frac{2\varepsilon}{\hbar}
        \sum^{}_{kf}
        \sum^{}_{hg}}
\times
        {1\over\hbar^2}
         \int_0^\infty  
%        \!  
        d\tau \, e^{-{\varepsilon\over\hbar}\tau}  
        \,  
         \int_0^\infty  
%        \!  
        d\tau' \, e^{-{\varepsilon\over\hbar}\tau'}  
        \,  
e^{-{i\over\hbar}\Delta E_q ({\bbox{p}}) \tau}
e^{+{i\over\hbar}\Delta E_{q'} ({\bbox{p}}') \tau'}
\langle
\rho_{q'}^{\scriptscriptstyle\dagger}\rho_q (\tau-\tau')
\rangle
\nonumber
\\
&&
-\frac{\varepsilon}{\hbar}
         \sum^{}_{p}
        \sum^{}_{qq'}{}'
        \tilde{t} (q) \tilde{t}^* (q') 
\{
\vert {\bbox{p}} \rangle         
\langle {\bbox{p}} \vert
,
\varrho
\}
\\
&&
\hphantom{\frac{2\varepsilon}{\hbar}
        \sum^{}_{kf}
        \sum^{}_{hg}}
\times
        {1\over\hbar^2}
         \int_0^\infty  
%        \!  
        d\tau \, e^{-{\varepsilon\over\hbar}\tau}  
        \,  
         \int_0^\infty  
%        \!  
        d\tau' \, e^{-{\varepsilon\over\hbar}\tau'}  
        \,  
e^{-{i\over\hbar}\Delta E_q ({\bbox{p}}) \tau}
e^{+{i\over\hbar}\Delta E_{q'} ({\bbox{p}}) \tau'}
\langle
\rho_{q'}^{\scriptscriptstyle\dagger}\rho_q (\tau-\tau')
\rangle
,  
\end{eqnarray*}
where 
$
        \left \langle  
\ldots
        \right \rangle
$
denotes the ensemble average over ${{\varrho}^{\text{m}}}$, $\rho_q
(t)$ the Heisenberg operator
$  
        e^{+{i\over\hbar} H_{\text{m}}t}  
        \rho_q  
        e^{-{i\over\hbar} H_{\text{m}}t}  
$, and the more compact notation $\Delta E_q ({\bbox{p}})=E_{p+q} - E_p$ for
the energy transfer has been used.
Since the system is supposed to be homogeneous, the correlation
function selects the contributions for which
${{\bbox{q}}}={{\bbox{q}}}'$, and exploiting the identity
\begin{displaymath}
  1=        \int dt \, \delta (t-[\tau'-\tau])
=
        \int dt \, 
        \int\! \frac{dE}{2\pi\hbar} \,
        e^{
        {
        i
        \over
         \hbar
        }
        E (t-[\tau'-\tau]) 
         }  
\end{displaymath}
we have
\begin{eqnarray*}
  {\cal L} (\varrho)=
&&
\frac{2\varepsilon}{\hbar}
         \sum^{}_{pp'}
        \sum^{}_{q}{}'
        |\tilde{t} (q)|^2
        e^{{i\over\hbar}{\bbox{q}}\cdot{\hat {{\sf x}}}}
        \vert {\bbox{p}} \rangle
        \langle {\bbox{p}} \vert
        {\hat \varrho}
        \vert {\bbox{p}}' \rangle
        \langle {\bbox{p}}' \vert
        e^{-{i\over\hbar}{\bbox{q}}\cdot{\hat {{\sf x}}}}
\\
&&
\hphantom{\frac{2\varepsilon}{\hbar}
        \sum^{}_{kf}
        \sum^{}_{hg}}
\times
        {1\over\hbar^2}
         \int_0^\infty  
        \!  
        d\tau \, e^{-{\varepsilon\over\hbar}\tau}  
        \,  
         \int_0^\infty  
        \!  
        d\tau' \, e^{-{\varepsilon\over\hbar}\tau'}  
        \,  
\int dE         \,
e^{-{i\over\hbar}[\Delta E_q ({\bbox{p}}) - E] \tau}
e^{+{i\over\hbar}[\Delta E_{q} ({\bbox{p}}') -E] \tau'}
\\
&&
\hphantom{
\frac{2\varepsilon}{\hbar}
        \sum^{}_{kf}
        \sum^{}_{hg}
\times
}
\times        {  
        1  
        \over  
         2\pi\hbar
        }  
        \int dt \, 
        e^{
        {
        i
        \over
         \hbar
        }
        E t
        }  
\langle
\rho_{q}^{\scriptscriptstyle\dagger}\rho_q (t)
\rangle
\nonumber
\\
&&
-\frac{\varepsilon}{\hbar}
         \sum^{}_{p}
        \sum^{}_{q}{}'
        |\tilde{t} (q)|^2
\{
\vert {\bbox{p}} \rangle         
\langle {\bbox{p}} \vert
,
\varrho
\}
\\
&&
\hphantom{\frac{2\varepsilon}{\hbar}
        \sum^{}_{kf}
        \sum^{}_{hg}
}
\times
        {1\over\hbar^2}
         \int_0^\infty  
        \!  
        d\tau \, e^{-{\varepsilon\over\hbar}\tau}  
        \,  
         \int_0^\infty  
        \!  
        d\tau' \, e^{-{\varepsilon\over\hbar}\tau'}  
        \,  
\int dE         \,
e^{-{i\over\hbar}[\Delta E_q ({\bbox{p}}) -E] \tau}
e^{+{i\over\hbar}[\Delta E_{q} ({\bbox{p}})-E] \tau'}
\\
&&
\hphantom{\frac{2\varepsilon}{\hbar}
        \sum^{}_{kf}
        \sum^{}_{hg}
\times}
\times        {  
        1  
        \over  
         2\pi\hbar
        }  
        \int dt \, 
        e^{
        {
        i
        \over
         \hbar
        }
        E t
        }  
\langle
\rho_{q}^{\scriptscriptstyle\dagger}\rho_q (t)
\rangle
.  
\end{eqnarray*}
We can now meaningfully undo the Laplace transform, coming to
\begin{eqnarray*}
  {\cal L} (\varrho)=
&&
\frac{2\varepsilon}{\hbar}
         \sum^{}_{pp'}
        \sum^{}_{q}{}'
        |\tilde{t} (q)|^2
        e^{{i\over\hbar}{\bbox{q}}\cdot{\hat {{\sf x}}}}
        \vert {\bbox{p}} \rangle
        \langle {\bbox{p}} \vert
        {\hat \varrho}
        \vert {\bbox{p}}' \rangle
        \langle {\bbox{p}}' \vert
        e^{-{i\over\hbar}{\bbox{q}}\cdot{\hat {{\sf x}}}}
\\
&&
\hphantom{\frac{2\varepsilon}{\hbar}
        \sum^{}_{kf}
        \sum^{}_{hg}}
\times
\int dE         \,
\frac{\varepsilon}{\pi}
\frac{1}{E-\Delta E_q ({\bbox{p}}) +i\varepsilon}
\frac{1}{E-\Delta E_q ({\bbox{p}}') -i\varepsilon}
\\
&&
\hphantom{\frac{2\varepsilon}{\hbar}
        \sum^{}_{kf}
        \sum^{}_{hg}\times}
\times
        {  
        1  
        \over  
         2\pi\hbar
        }  
        \int dt \, 
        e^{
        {
        i
        \over
         \hbar
        }
        E t
        }  
\langle
\rho_{q}^{\scriptscriptstyle\dagger}\rho_q (t)
\rangle
\nonumber
\\
&&
-\frac{\varepsilon}{\hbar}
         \sum^{}_{p}
        \sum^{}_{q}{}'
        |\tilde{t} (q)|^2
\{
\vert {\bbox{p}} \rangle         
\langle {\bbox{p}} \vert
,
\varrho
\}
\\
&&
\hphantom{\frac{2\varepsilon}{\hbar}
        \sum^{}_{kf}
        \sum^{}_{hg}}
\times
\int dE         \,
\frac{\varepsilon}{\pi}
\frac{1}{E-\Delta E_q ({\bbox{p}}) -i\varepsilon}
\frac{1}{E-\Delta E_q ({\bbox{p}}) +i\varepsilon}
 \\
&&
\hphantom{\frac{2\varepsilon}{\hbar}
        \sum^{}_{kf}
        \sum^{}_{hg}\times}
\times
       {  
        1  
        \over  
         2\pi\hbar
        }  
        \int dt \, 
        e^{
        {
        i
        \over
         \hbar
        }
        E t
        }  
\langle
\rho_{q}^{\scriptscriptstyle\dagger}\rho_q (t)
\rangle
.  
\end{eqnarray*}
If we now exploit the quasi-diagonality of $\varrho$, linked to its
slow variability, thus substituting in the denominators of the first
term
${\bbox{p}}$, ${\bbox{p}}'$
with the symmetric expression $\frac 12 ({\bbox{p}} +
{\bbox{p}}')$, we obtain the expression 
\begin{eqnarray}
  \label{pra3}
  {\cal L} (\varrho)=
&&
\frac{2\pi}{\hbar}
         \sum^{}_{pp'}
        \sum^{}_{q}{}'
        |\tilde{t} (q)|^2
        e^{{i\over\hbar}{\bbox{q}}\cdot{\hat {{\sf x}}}}
        \vert {\bbox{p}} \rangle
        \langle {\bbox{p}} \vert
        {\hat \varrho}
        \vert {\bbox{p}}' \rangle
        \langle {\bbox{p}}' \vert
        e^{-{i\over\hbar}{\bbox{q}}\cdot{\hat {{\sf x}}}}
        {  
        1  
        \over  
         2\pi\hbar
        }  
        \int dt \, 
        e^{
        {
        i
        \over
         \hbar
        }
        \Delta E_q (\frac {{\bbox{p}} +
{\bbox{p}}'}{2} )t
        }  
\langle
\rho_{q}^{\scriptscriptstyle\dagger}\rho_q (t)
\rangle
\\
&&
\nonumber
-\frac{\pi}{\hbar}
         \sum^{}_{p}
        \sum^{}_{q}{}'
        |\tilde{t} (q)|^2
\{
\vert {\bbox{p}} \rangle         
\langle {\bbox{p}} \vert
,
\varrho
\}
        {  
        1  
        \over  
         2\pi\hbar
        }  
        \int dt \, 
        e^{
        {
        i
        \over
         \hbar
        }
        \Delta E_q ({\bbox{p}}) t
        }  
\langle
\rho_{q}^{\scriptscriptstyle\dagger}\rho_q (t)
\rangle
.    
\end{eqnarray}
The correlation functions appearing in (\ref{pra3}) are exactly the
dynamic
structure factor multiplied by $N$ and evaluated for a momentum transfer
${\bbox{q}}$ and energy transfers $\Delta E_q (\frac {{\bbox{p}} +
{\bbox{p}}'}{2} )$ and $\Delta E_q ({\bbox{p}})$ respectively, 
as can be seen by comparison with (\ref{qq}).
To see under which conditions the obtained master equation (\ref{xx})
takes a Lindblad structure we consider an approximation of the form
(\ref{x}), which will generally depend on the smoothness of the energy 
dependence of the dynamic
structure factor, but is actually less demanding than it might
seem, since in the expression (\ref{pra3}) one has to consider a sum
over ${\bbox{p}}$ and ${\bbox{p}}'$ with the matrix elements 
$\langle {\bbox{p}} \vert
        {\hat \varrho}
        \vert {\bbox{p}}' \rangle
$ 
of the statistical operator.
In the continuum limit we therefore obtain the master equation
\begin{eqnarray}
  \label{general}
        {  
        d {\hat \varrho}  
        \over  
                      dt
        }  
        &=&
        -
        {i \over \hbar}
        [
        {\hat {{\sf H}}}_0
        ,
        {\hat \varrho}
        ]
        +
        {\cal L} (\varrho)
        \\
        &=&
        -
        {i \over \hbar}
        \left[
        {
        {\hat {{\sf p}}}^2
        \over
        2M
        }
        ,
        {\hat \varrho}
        \right]
        +
        {2\pi \over\hbar}
        (2\pi\hbar)^3
        n
        \int d^3\!
        {\bbox{q}}
        \,  
        {
        | \tilde{t} (q) |^2
        }
        \Biggl[
        e^{{i\over\hbar}{\bbox{q}}\cdot{\hat {{\sf x}}}}
        \sqrt{
        S({\bbox{q}},{\hat {{\sf p}}})
        }
        {\hat \varrho}
        \sqrt{
        S({\bbox{q}},{\hat {{\sf p}}})
        }
        e^{-{i\over\hbar}{\bbox{q}}\cdot{\hat {{\sf x}}}}
        -
        \frac 12
        \left \{
        S({\bbox{q}},{\hat {{\sf p}}}),
        {\hat \varrho}
        \right \}
        \Biggr],
        \nonumber
\end{eqnarray}
which still has the form  (\ref{ss}), but is much more general since 
now the dynamic structure factor
does not necessarily describe a free gas.
This result allows for the extension of the usefulness of
the master equation to cases in which the correlation
function cannot be directly evaluated, but a suitable
phenomenological model is available, e.g., determined
in terms of scattering experiments.
%%%%%%%%%%%%%%%%%%%%%%%%%%%%%%%%%%%%%%%%%%%%%%%%%%%%%%%%%%%%%
\section{QUANTUM BROWNIAN MOTION AND QUANTUM  STATISTICS}
%%%%%%%%%%%%%%%%%%%%%%%%%%%%%%%%%%%%%%%%%%%%%%%%%%%%%%%%%%%%%
We are now interested in 
the Brownian limit
$\alpha=m/M\ll 1$, considering the dynamics of a free particle
interacting through collisions with a gas of much lighter
particles. Having an expression valid for both a Fermi or
Bose gas it is particularly interesting to evaluate the
correction brought about by quantum statistics to the
typical models of quantum Brownian motion.
In the  limit
$\alpha\ll 1$ expressions (\ref{7}) and (\ref{8}) become respectively
        \begin{equation}
        \label{10}
%        &&
        S_{\rm \scriptscriptstyle B/F}({\bbox{q}},{\bbox{p}},\alpha\ll 1)
        =
        {
        1
        \over
         (2\pi\hbar)^3
        }
        {
        2\pi m^2
        \over
        n\beta q
        }
%        \\
%        &&
%        \hphantom{\times\times}
%        \times
        {
        \mp
        \over
        1-
        e^{
        {
        \beta
        }
        \left[
        \frac{q^2}{2M}
        +\frac{{\bbox{q}}\cdot{\bbox{p}}}{M}
        \right]
        }
        }
        \log
        \left[
        {
        1\mp z
        e^{
        -{
        \beta
        \over
             8m
        }
        q^2
        }
        e^{
        -\frac{\beta}{2}
        [
        {
        q^2
        \over
             2M
        }
        +
        {{\bbox{q}}\cdot{\bbox{p}} \over M}        
        ]
        }
        \over
        1\mp z
        e^{
        -{
        \beta
        \over
             8m
        }
        q^2
        }
        e^{
        +\frac{\beta}{2}
        [
        {
        q^2
        \over
             2M
        }
        +
        {{\bbox{q}}\cdot{\bbox{p}} \over M}
        ]
        }
        }
        \right]
%        \nonumber
        \end{equation}
        \begin{equation}
        \label{9}
        S_{\rm \scriptscriptstyle MB}({\bbox{q}},{\bbox{p}},\alpha\ll 1)
        =
        {
        1
        \over
         (2\pi\hbar)^3
        }
        {
        2\pi m^2
        \over
        n\beta q
        }
        z
        e^{
        -{
        \beta
        \over
             8m
        }
        q^2
        }
        e^{
        -\frac{\beta}{2}
        [
        {
        q^2
        \over
             2M
        }
        +
        {{\bbox{q}}\cdot{\bbox{p}} \over M}
        ]
        }
        ,
        \end{equation}
or expressed in terms of momentum and energy transfer
        \begin{eqnarray*}
        S_{\rm \scriptscriptstyle B/F}(q,E,\alpha\ll 1)
        &=&
        {
        1
        \over
         (2\pi\hbar)^3
        }
        {
        2\pi m^2
        \over
        n\beta q
        }
%        \\
%        &&
%        \hphantom{\times\times}
%        \times
        {
        \mp
        \over
        1-
        e^{\beta E
        }
        }
        \log
        \left[
        {
        1\mp z
        e^{
        -{
        \beta
        \over
             8m
        }
        q^2
        }
        e^{
        -\frac{\beta}{2}
        E
        }
        \over
        1\mp z
        e^{
        -{
        \beta
        \over
             8m
        }
        q^2
        }
        e^{
        +\frac{\beta}{2}
        E
        }
        }
        \right]
        \\
        S_{\rm \scriptscriptstyle MB}(q,E,\alpha\ll 1)
        &=&
        {
        1
        \over
         (2\pi\hbar)^3
        }
        {
        2\pi m^2
        \over
        n\beta q
        }
        z
        e^{
        -{
        \beta
        \over
             8m
        }
        q^2
        }
        e^{
        -\frac{\beta}{2}
        E
        }
        ,
        \end{eqnarray*}
still satisfying the principle of detailed balance~\cite{Lovesey}.
In the Boltzmann case, as mentioned above expression  (\ref{9}) exactly 
fulfills  (\ref{x}) and 
the
generator in (\ref{6}) takes the particularly simple form
$
        L_{\rm \scriptscriptstyle B/F}({\bbox{q}},{\hat {{\sf p}}},{\hat
        {{\sf x}}})
        \propto
        e^{{i\over\hbar}{\bbox{q}}\cdot{\hat {{\sf x}}}}
        e^{
        -{
        \beta
        \over
             4M
        }
        {\bbox{q}}\cdot{\hat {{\sf p}}}
        }
$,
so that one obtains for an isotropic medium the master
equation given in~\cite{art3}
        \begin{eqnarray}
        \label{also}
        {  
        d {\hat \varrho}  
        \over  
                dt  
        }  
        =
        &-&
        {i\over\hbar}
        [
        {{\hat {\mbox{\sf H}}}_0}
        ,{\hat \varrho}
        ]
        \\
        &+&
        z
        {4\pi^2 m^2 \over\beta\hbar}
        \int d^3\!
        {\bbox{q}}
        \,  
        {
        | \tilde{t} (q) |^2
        \over
        q
        }
        e^{-
        {
        \beta
        \over
             8m
        }
        {{{q}}^2}
        }
        \Biggl[
        e^{{i\over\hbar}{\bbox{q}}\cdot{\hat {{\sf x}}}}
        e^{-{\beta\over 4M}{\bbox{q}}\cdot{\hat {{\sf p}}}}
        {\hat \varrho}
        e^{-{\beta\over 4M}{\bbox{q}}\cdot{\hat {{\sf p}}}}
        e^{-{i\over\hbar}{\bbox{q}}\cdot{\hat {{\sf x}}}}
        - {1\over 2}
        \left \{
        e^{-{\beta\over 2M}{\bbox{q}}\cdot{\hat {{\sf p}}}}
        ,
        {\hat \varrho}
        \right \}
        \Biggr]
        .
        \nonumber
        \end{eqnarray}
To recover the equation describing quantum Brownian motion one goes over to
small momentum transfer, strongly favored by the kinematics
of the collisions, considering terms up to second order in ${\bbox{q}}$
or equivalently
bilinear in
${\hat {{\sf x}}}$ and
${\hat {{\sf p}}}$, thus obtaining an equation in close
analogy to the classical  description, with a friction force
proportional to velocity.
The result for a Boltzmann gas is
        \begin{eqnarray}
        \label{boltz}
        {  
        d {\hat \varrho}  
        \over  
                dt  
        }  
        &=&
        -
        {i\over\hbar}
        [
        {{\hat {\mbox{\sf H}}}_0}
        ,{\hat \varrho}
        ]
        -
        z
        \sum_{i=1}^3
        \left \{
        {
        D_{pp}
        \over
         \hbar^2
        }
        \left[  
        {\hat {{\sf x}}}_i,
        \left[  
        {\hat {{\sf x}}}_i,{\hat \varrho}
        \right]  
        \right]
        \right.
        \nonumber 
        \\
        &&
        \hphantom{\quad\quad}
        +
        \left.
        {
        D_{xx}
        \over
         \hbar^2
        }
        \left[  
        {\hat {\mbox{\sf p}}}_i,
        \left[  
        {\hat {\mbox{\sf p}}}_i,{\hat \varrho}
        \right]  
        \right]  
        +{i\over\hbar}
        \gamma
        \left[  
        {\hat {{\sf x}}}_i ,
        \left \{  
        {\hat {\mbox{\sf p}}}_i,{\hat \varrho}
        \right \}  
        \right]
        \right \}
        \end{eqnarray}
with
        \begin{eqnarray}
        \label{gamma}  
        D_{pp}
        &=&
        \frac 23
        {\pi^2 m^2 \over\beta\hbar}
        \int d^3\!
        {\bbox{q}}
        \,  
        {
        | \tilde{t} (q) |^2
        }
        q
        e^{-
        {
        \beta
        \over
             8m
        }
        {{{q}}^2}
        },
        \nonumber
        \\
        D_{xx} 
        &=& 
        (
        {
        \beta\hbar
        /
            4M
        })^2 D_{pp},
        \quad
        \gamma =
        ({
        \beta
        /
             2M
        })
        D_{pp},
        \end{eqnarray}        
and has the particular feature that it can be written in
Lindblad form in terms of a single generator~\cite{art3}.
Starting from (\ref{10}) one can perform the same limit of
small momentum transfer corresponding through the physical
interpretation of the dynamic
structure factor to the macroscopic, long
wavelength properties of the system, thus calculating the
correction due to quantum statistics to the master equation
describing quantum Brownian motion. 
To do this one considers the Taylor expansion of the logarithms in
(\ref{10}), leading to the following compact expression as a power
series in the fugacity $z$
\begin{eqnarray}
  &&
  \label{serie}
  S_{\rm \scriptscriptstyle B/F}(q,E,\alpha\ll 1)= 
  S_{\rm
    \scriptscriptstyle MB}(q,E,\alpha\ll 1)
  \\
  &&
  \hphantom{\times\times\times}
  \times
  \left[
    1+\sum_{k=1}^{\infty}{(\pm)}^k
    \frac{z^k}{k+1}e^{-k\frac{\beta}{8m}q^2}
    e^{-k\frac{\beta}{2}E} \sum_{n=0}^{k}e^{n \beta E}
  \right],  
  \nonumber
\end{eqnarray}
which has to be substituted in (\ref{ss}), keeping terms up to second
order in $q$. 
The result one obtains
is actually remarkably simple: 
the operator structure is not changed, nor the
simple generator feature, but the fugacity appears through
the expression
$
{
z
/ 
(1\mp z)
}$
rather than linearly. For a Bose or Fermi gas at finite
temperature one has
        \begin{eqnarray}
        \label{11}
        {  
        d {\hat \varrho}  
        \over  
                dt  
        }  
        &=&
        -
        {i\over\hbar}
        [
        {{\hat {\mbox{\sf H}}}_0}
        ,{\hat \varrho}
        ]
        -
        {
        z
        \over
         1\mp z
        }
        \sum_{i=1}^3
        \left \{
        {
        D_{pp}
        \over
         \hbar^2
        }
        \left[  
        {\hat {{\sf x}}}_i,
        \left[  
        {\hat {{\sf x}}}_i,{\hat \varrho}
        \right]  
        \right]
        \right.
        \nonumber 
        \\
        &&
        \hphantom{\quad\quad}
        +
        \left.
        {
        D_{xx}
        \over
         \hbar^2
        }
        \left[  
        {\hat {\mbox{\sf p}}}_i,
        \left[  
        {\hat {\mbox{\sf p}}}_i,{\hat \varrho}
        \right]  
        \right]  
        +{i\over\hbar}
        \gamma
        \left[  
        {\hat {{\sf x}}}_i ,
        \left \{  
        {\hat {\mbox{\sf p}}}_i,{\hat \varrho}
        \right \}  
        \right]
        \right \}
        \end{eqnarray}
where the coefficient
$
        {
        z
        /
         (1\mp z)
        }
$
at finite temperature is actually well defined because $z$
is in the range $0\leq z < 1$ for Bose particles and
positive for Fermi particles.
\par
Eq.~(\ref{11}) expressing the correction due to quantum
statistics in the equation describing quantum Brownian motion, 
together with Eq.~(\ref{general}) 
giving  a completely positive time evolution for a
particle interacting with a macroscopic system at
equilibrium in terms of a momentum displacement operator and
the dynamic
structure factor of the system, are the main results of this paper.
Comparing (\ref{11}) with (\ref{boltz}) one sees that the
friction coefficient given in the Boltzmann case by 
        \begin{equation}
        \label{f1}
        \gamma_{\rm \scriptscriptstyle MB}=
        z
        {
        \beta
        \over
             2M
        }
        D_{pp}
        =
        z
        \frac 13
        {\pi^2 m^2 \over M\hbar}
        \int d^3\!
        {\bbox{q}}
        \,  
        {
        | \tilde{t} (q) |^2
        }
        q
        e^{-
        {
        \beta
        \over
             8m
        }
        {{{q}}^2}
        }
        \end{equation}
is now substituted by
\begin{equation}
        \label{f2}
        \gamma_{\rm \scriptscriptstyle B/F}=
        {
        \gamma_{\rm \scriptscriptstyle MB}
        \over
         1\mp z
        }
\end{equation}
enhanced or suppressed according to statistics.
Both (\ref{boltz}) and (\ref{11}) retain the
property of complete positivity satisfied by (\ref{ll}), are invariant under 
translation and rotation and admit a stationary solution of the
canonical form
$
{\hat \varrho} \propto e^{-
\beta
{
{\hat {{\sf p}}}^2
\over
     2M
}
}
$.
The single generator feature is due to the fact that the
coefficients satisfy the relationship
$
D_{pp}D_{xx} =
{\hbar^2 \gamma^2 / 4}
$.
%%%%%%%%%%%%%%%%%%%%%%%%%%%%%%%%%%%%%%%%%%%%%%%%%%%%%%%%%%%%%
\section{SUMMARY AND OUTLOOK}  
%%%%%%%%%%%%%%%%%%%%%%%%%%%%%%%%%%%%%%%%%%%%%%%%%%%%%%%%%%%%%
We have considered the problem of the motion of a test particle
interacting through collisions with a fluid, following the approach
outlined in~\cite{art1,berlin,japan}, which has already been successfully
applied to the case of neutron optics~\cite{art2}.
The microscopic derivation allows some insights into the conditions
under which a master equation of the Lindblad type, driving a 
completely positive time
evolution, can be obtained, thus giving a concrete physical example
contributing to the debate on the relevance of
complete positivity~\cite{Pechukas-Piza}. 
Provided the statistical operator is
sufficiently diagonal in momentum representation with respect to the
energy dependence of the dynamic
structure factor, the master equation (\ref{general}) is
obtained, where only quantities of  physical
interest appear: the scattering cross section for
the single two-body collisions, given by the square of the
T matrix; the generator of translations in momentum space
and the dynamic
structure factor, keeping the statistical properties of the
medium into account, combined through the expression 
$        L({\bbox{q}},{\hat {{\sf p}}},{\hat
        {{\sf x}}})
        =
        e^{{i\over\hbar}{\bbox{q}}\cdot{\hat {{\sf x}}}}
        \sqrt{
        S({\bbox{q}},{\hat {{\sf p}}})
        }
$. 
This structure is remarkably simple and describes a dynamics in which
the motion of the test particle is linked through this particular
two-point correlation function to the spectrum of spontaneous
fluctuations of the system.
Starting from this general structure and explicitly calculating the
dynamic
structure factor for the case of a free gas 
one can consider the particularly
relevant case of Brownian motion, when the test particle is much
heavier than the particles making up the gas.
In the case of a Boltzmann gas one recovers, for small momentum
transfer, a typical structure of generator of quantum 
Brownian motion, given by Eq.~(\ref{boltz}), 
in which all coefficients are determined and the
dissipative part of the generator depends linearly on the fugacity. 
The case of a quantum gas is also considered, and in this case the
generator has the structure (\ref{11}), with the dissipative part
depending on the fugacity $z$ through the expression ${z}/ ({1\mp
  z})$, thus giving the connection (\ref{f2}) between the friction
coefficients in the different cases.
\par
We hope that this fundamental study on the general features of a
master equation describing the motion of a test particle in a gas,
putting in major evidence the dynamic
structure factor and showing the relationship
between this structure and the equation, analogous to the
Fokker-Planck equation, describing quantum Brownian motion, could be a sound
starting point for future extensions and applications,
especially in connection with degenerate regimes at very low
temperatures, where the dynamic
structure factor is now being intensively studied both at
experimental~\cite{ketterle-s} and theoretical level~\cite{stringari}.
%%%%%%%%%%%%%%%%%%%%%%%%%%%%%%%%%%%%%%%%%%%%%%%%%%%%%%%%%%%%%
\section*{ACKNOWLEDGMENTS}
%%%%%%%%%%%%%%%%%%%%%%%%%%%%%%%%%%%%%%%%%%%%%%%%%%%%%%%%%%%%%
The author would like to thank Prof. L. Lanz for many useful
suggestions and careful reading of the manuscript and Prof. A.
Barchielli for useful discussions. This work was partially supported
by the Alexander von Humboldt-Stiftung and by MURST under
Cofinanziamento and Progetto Giovani.
%%%%%%%%%%%%%%%%%%%%%%%%%%%%%%%%%%%%%%%%%%%%%%%%%%%%%%%%%%%%%
  

\begin{references}  

\bibitem{Kiefer}
{D. Giulini~{\it et al.}},
{\it Decoherence and the Appearance of a Classical World in
Quantum Theory}
(Springer, Berlin, 1996).


\bibitem{qed-ion}
{M.~Brune~{\it et al.}},
{Phys.~Rev.~Lett.}
{\bf 77},
4887
(1996);
{C.~J.~Myatt~{\it et al.}},
{Nature}
{\bf 403},
269
(2000).


\bibitem{bec}
{M.~Anderson~{\it et al.}},
{Science}
{\bf 269},
198
(1995); 
{F.~Dalfovo~{\it et al.}},
{Rev.~Mod.~Phys.}
{\bf 71},
463
(1999);
{G.~M.~Tino and M.~Inguscio},
{Rivista del Nuovo Cimento}
{\bf 22},
1 
(1999).


\bibitem{fermi}
{B.~DeMarco and D.~S.~Jin},
{Science}
{\bf 285},
1703
(1999).



\bibitem{Imoto-Ferrari-QKI}
{M.~Yamashita, M.~Koashi and N.~Imoto},
{Phys.~Rev.~A}  
{\bf 2243},
{59}
(1999);
{G.~Ferrari},
{Phys.~Rev.~A}  
{\bf 59},
{R4125}
(1999);
{C.~W.~Gardiner and P.~Zoller},  
{Phys.~Rev.~A}  
{\bf 55},  
{2902}  
({1997}).
  

\bibitem{art3}
{B.~Vacchini},
{Phys.~Rev.~Lett.}
{\bf 84},
1374
(2000).


\bibitem{LindbladQBM-AlbertoQBM-LindbladJMP}
G.~Lindblad,
{Rep.~Math.~Phys.}
{\bf 10},
393
(1976);
A.~Barchielli,
{Nuovo Cimento}
{\bf 74B},
113
(1983);
{G.~Lindblad},  
{J.~Math.~Phys.}
{\bf 39},  
{2763}  
({1998}).  
 

\bibitem{Lindblad-Kraus-Hellwig-Alicki}
{G.~Lindblad},  
{Commun.~Math.~Phys.}
{\bf 48},  
{119}  
({1976});
K.~Kraus,
{\it Lect.~Notes in Physics}, Vol. 190
({Springer}, {Berlin}, 1983);
K.-E.~Hellwig,
{Int. J. Theor. Phys.}
{\bf 34},
{1467}
(1995);
{R.~Alicki and K.~Lendi},
{\it Lect.~Notes in Physics}, Vol.~286
(Springer, Berlin, 1987).


\bibitem{Sandulescu-Isar}
{A.~S\v{a}ndulescu and H.~Scutaru},
{Ann.~Phys. (N.~Y.)}
{\bf 173},
277
(1987);
{A.~Isar},  
{Fortschr.~Phys.}
{\bf 47},  
{855}  
({1999}).  
 


\bibitem{Ambegaokar-Pechukas2-IsarJMP-Tannor-97}
V.~Ambegaokar,
{\it Ber.~Bunsenges~Phys.~Chem.}
{\bf 95},
400
(1991);
P.~Pechukas, in {\it Large Scale Molecular Systems}, edited
by W.~Gans, A.~Blumen, and A.~Amann, NATO ASI Series (Plenum
Press, New York, 1991), Vol.~258, p.~123;
{A.~Isar, A.~Sandulescu and W.~Scheid},  
{J.~Math.~Phys.}
{\bf 34},  
{3887}  
({1993});
D.~Kohen, C.~C.~Marston and D.~J.~Tannor,
{J.~Chem.~Phys.}
{\bf 107},
5236
(1997).  

\bibitem{Pechukas-Piza}
{P.~Pechukas},
{Phys.~Rev.~Lett.}
{\bf 73},
1060
(1994);
{\bf 75},
3021
(1995);
R.~Alicki, {\it ibid.}
{\bf 75},
3020
(1995);
{R.~C.~de~Berr\^{e}do~{\it et al.}},
{Physica Scripta}
{\bf 57},
533
(1998).


\bibitem{art1}  
L.~Lanz and B.~Vacchini,  
{Int. J. Theor. Phys.}
{\bf 36},  
67  
(1997).
  

\bibitem{vanHove}  
{L.~van~Hove},  
{ Phys.~Rev.}  
{\bf 95},  
{249}  
(1954).  
  

\bibitem{Lovesey}  
{S.~W.~Lovesey},  
{\it Theory of Neutron Scattering from Condensed Matter}  
(Clarendon Press, Oxford, 1984).  
 


\bibitem{Glyde}  
{H.~R.~Glyde},  
{\it Excitations in Solid and Liquid Helium}  
(Clarendon Press, Oxford, 1994).  
   
\bibitem{berlin}  
L.~Lanz and B.~Vacchini,  
{ Int. J. Theor. Phys.}
{\bf 37},  
545  
(1998).  
  

\bibitem{japan}  
L.~Lanz, O.~Melsheimer and B.~Vacchini, in  
{\it Quantum communication, computing, and measurement}, edited by   
O.~Hirota, A.~S.~Holevo and C.~M.~Caves, (Plenum, New York, 1997),  
p.~339.

\bibitem{art2}
L.~Lanz and B.~Vacchini,  
{Phys. Rev. A}
{\bf 56},
4826
(1997).

\bibitem{ketterle-s}
{J.~Stenger~{\it et al.}},
{Phys.~Rev.~Lett.}
{\bf 82},
4569
(1999);
{D.~M.~Stamper-Kurn~{\it et al.}},
{Phys.~Rev.~Lett.}
{\bf 83},
2876
(1999).

\bibitem{stringari}
F.~Zambelli~{\it et al.},
{Phys. Rev. A}
{\bf 61},
063608
(2000).

\end{references}
\end{document}